\documentclass[prd,showpacs]{revtex4}
\usepackage[dvips]{graphicx,psfrag}
\usepackage{amssymb}
\input epsf
\newcommand*{\be}{\begin{equation}}
\newcommand*{\ee}{\end{equation}}
\newcommand*{\bea}{\begin{eqnarray}}
\newcommand*{\eea}{\end{eqnarray}}
\newcommand*{\lb}{\label}
\begin{document}
\title{ACCURATE RESULTS FOR PRIMORDIAL BLACK HOLES 
        FROM SPECTRA WITH A DISTINGUISHED SCALE}
\author{David Blais}
\affiliation{
Laboratoire de Physique Math\'ematique et Th\'eorique,
 UMR 5825 CNRS,\\
 Universit\'e de Montpellier II, 34095 Montpellier, France.}
\author{Torsten Bringmann}
\affiliation{
 Department of Physics, Stockholm University,
  10691 Stockholm, Sweden.}
\author{Claus Kiefer}
\affiliation{
 Institut f\"ur Theoretische Physik, Universit\"at zu K\"oln,
 Z\"ulpicher Str.~77, 50937 K\"oln, Germany.}
\author{David Polarski}
\affiliation{ Laboratoire de Physique Math\'ematique et Th\'eorique,
 UMR 5825 CNRS,
 Universit\'e de Montpellier II, 34095 Montpellier, France\footnote{permanent
address after September 1, 2002.}}
\affiliation{Laboratoire de Math\'ematiques et Physique Th\'eorique,
 UMR 6083 CNRS,
  Universit\'e de Tours, Parc de Grandmont, 37200 Tours, France.}

\date{\today}
\begin{abstract}
We perform an accurate computation of the production rate for
primordial black holes (PBHs). The reason is that the underlying 
mass variance had been overestimated systematically, as was
shown recently.
For scale-free powerlaw primordial spectra, and for a Universe with
critical density,  
the mass variance is less than 34\% of its value thought earlier for 
the spectral index in the range
$1\leq n\leq 1.3$. We then extend our study to spectra with 
a characteristic scale and find the accurate shape of the 
corresponding mass variance. 
For a pure step in the primordial spectrum, 
the step in the variance is smoothed around the 
characteristic scale $k_s$. For a spectrum with large oscillations 
near $k_s$, we find 
a pronounced bump in the variance. This could yield 
a significant part of the cold dark matter in the 
form of PBHs with mass $M$ in the 
range $5\times 10^{15}\ {\rm g}\lesssim M \lesssim 10^{21}\ {\rm g}$.
\end{abstract}
\pacs{04.62.+v, 98.80.Cq}
\maketitle

\section{Introduction}

Primordial black holes (PBHs) are black holes that result from the
collapse of density fluctuations in the early Universe \cite{CH74}.
 This is the only way
to produce black holes with mass smaller than about three solar masses
(except possibly by the production of black holes in accelerators,
as has recently been discussed in the speculative framework of branes).
This makes them interesting for several reasons. First, PBHs with initial
mass $\lesssim M_*\approx 5\times 10^{14}\ {\rm g}$ have been already
evaporated by Hawking radiation or evaporate at the present stage
of the Universe. They would play a key role in studying quantum
gravitational effects of black holes (see, e.g., \cite{GBH}).
Second, PBHs with any mass could be an excellent indicator of the
conditions in the early Universe. And third, they could in principle
contribute a major part to dark matter. As no PBHs have yet been observed
(or identified as such), existing observational limits can strongly
constrain scenarios of the early Universe.

One of the most successful scenarios is inflation (see, e.g., \cite{LL}
for a review). It leads to a primordial spectrum of fluctuations
that can serve as the seed for the observed structure in the Universe.
Recent observations of the anisotropies in the cosmic microwave
background (CMB) radiation strongly support inflation. 
Depending on the exact form of the spectrum, the same fluctuations can also
lead to the formation of PBHs (see, e.g., \cite{Carr2,GL}). Most models of
inflation predict a scale-free spectrum with a spectral index $n$
very close to the scale-invariant case $n=1$. A significant number of PBHs
can only be produced for $n>1$ (a ``blue spectrum''),
since these values lead to more power on small scales
(this was first addressed in \cite{CGL}). Observational
limits (both from Hawking radiation and the fact that PBHs must
not overclose the Universe) strongly constrain $n$. This, therefore,
yields a constraint on inflationary models that is independent of the
cosmological constraints from the CMB and large-scale structure.

It has been found recently that the production rate of PBHs
was systematically overestimated \cite{BKP1}. The reason is that one
has to use the transfer function (the tool to describe the evolution
of fluctuations within the horizon) and evaluate it at the onset
of PBH formation, not at the present time.
The qualitative explanation goes as follows: 
at each stage of the cosmological evolution, the correct use of the
transfer function takes into account the dynamics of the fluctuations
after horizon-entry. This leads effectively to very different spectra
on small scales at different times.
A general formula to remedy
this earlier deficiency has been presented in \cite{BKP1} and
some quantitative estimates have been given. The extension from vanishing
cosmological constant discussed in \cite{BKP1} to the nonvanishing case
has been performed in \cite{P02}. However, an accurate calculation
has to be done numerically. This will be done here and will constitute
the first aim of the present paper. 

Although scale-free models have been successful so far, it is not clear 
that the true spectrum must be exactly scale-free. The inflaton potential
may exhibit distinguished features at an energy scale relevant to
GUT theories. This would lead to models with broken scale invariance
(BSI). One simple example would be a jump in the derivative of the
inflaton potential, interrupting the slow-roll behavior and leading to
additional particle creation and formation of PBHs. The spectrum of
fluctuations resulting from such a jump can be presented analytically
in closed form \cite{S92}. It can cope surprisingly well with
observational constraints like the possible presence of a bump 
in the matter power spectrum \cite{LPS98}. The BSI model contains two
additional parameters: the strength of the jump, $p$, and its
location, $k_s$. 

In \cite{BKP1} we have already discussed the simple case of a
pure step in the mass variance. The BSI model of \cite{S92} is, of course,
of much more physical relevance. The second aim of our paper is thus
to perform an accurate calculation of PBH production in this model.
In fact, the production rate can be so high that the PBHs could be
candidates for massive compact halo objects (MACHOs) and thus contribute
a significant part of the dark matter in the Universe \cite{BKP2}.

BSI models have the advantage that they are the only realistic models
of PBH formation. The reason is that scale-free models need a certain
fine-tuning to lead to a significant number of PBHs: if the spectral index
is only a little more than some critical value $n^*$, the PBH production
rate would be so high that it were in conflict with observation.
On the other hand, for $n$ being somewhat smaller than $n^*$, the production
rate would be exponentially small and one would never have the
chance to observe a PBH. The critical value is $n^*\approx 1.3$ \cite{BKP1}
which is slightly bigger than current observational limits
(see e.g. \cite{Max01}). Even more, a blue spectrum with constant 
spectral index close to $n^*$ is inevitably ruled out due to the overproduction 
of PBHs which have evaporated by now and contribute to the diffuse $\gamma$-ray background. 
The situation will be drastically different in a BSI model,
as will become clear from the results presented here.
PBH production from models with a scale was considered before
\cite{INN94,RSG96,BP}, but we shall reconsider it here in the light of our
improved formulae from \cite{BKP1}.

Our paper is organized as follows. In Sec.~II we shall briefly review
the fundamental formalism to calculate PBH formation,
see \cite{CH74} and \cite{BKP1} for more details. In particular,
we shall present the crucial relation between primordial power spectrum
and mass variance. In Sec.~III we then give accurate results for
the cases of scale-free spectra, spectra with a pure step, and the
BSI model of \cite{S92}. Sec.~IV is devoted to the discussion of
observational constraints for each of these situations. For the
BSI model, in particular, these constraints allow the possibility to
form PBHs in the mass range 
$5\times 10^{15}\ {\rm g}\lesssim M \lesssim 10^{21}\ {\rm g}$. Such
PBHs could play the role of MACHOs in the halos of galaxies and could
contribute significantly to the dark matter in the Universe.

\section{PBH formation}

It was realized some time ago that a spectrum of 
primordial density fluctuations would inevitably 
lead to the production of primordial black holes.
Let us assume for simplicity that a PBH is formed 
when the density contrast averaged over a volume 
of the (linear) size of the Hubble radius satisfies 
$\delta_{min} \leq \delta \leq \delta_{max}$,
where the PBH mass $M_{PBH}$ is just the ``horizon mass'' $M_H$, 
the mass contained inside the Hubble
volume. Usually one takes 
$\delta_{min}\approx\frac{1}{3}$ and $~\delta_{max}=1$; 
however, recent numerical simulations suggest
a higher value of $\delta_{min}\approx 0.7$ \cite{NJ98}.

Each physical scale $\lambda(t)$ is defined by some wavenumber $k$ and 
evolves with time according to $\lambda(t)= 2\pi a(t)/k$. 
For a given physical scale, the ``horizon'' crossing time $t_k$ 
is conventionally defined by 
$k=a(t_k)H(t_k)$. It is the time when that scale re-enters the Hubble radius 
(here we do not mean 
``horizon'' crossing during inflation but after inflation), which will 
inevitably happen after 
inflation for scales that are larger than the Hubble radius at 
the end of inflation. 
It is at that time $t_k$ that a PBH might form with mass 
$M_H(t_k)$. Clearly, there is a one-to-one correspondence between 
$\lambda(t_k), M_H(t_k)$, and $k$. 
We can of course also take this correspondence at any other 
initial time $t_i$ and relate the physical 
quantities at both times $t_i$ and $t_k$. 

Generally, if the primordial fluctuations obey a Gaussian 
statistics, the probability density $p(\delta)$, where $\delta$ is the 
density contrast averaged over 
a sphere of radius $R$, is given by
\be
p(\delta) = \frac{1}{\sqrt{2\pi}~\sigma (R)}~ 
e^{-\frac{\delta^2}{2 \sigma^2(R)}},
\ee 
where the dispersion $\sigma^2(R) \equiv \Bigl \langle \Bigl ( \frac{\delta M}{M}  
\Bigr )_R^2 \Bigr \rangle$ is computed using a top-hat window function, 
\be
\sigma^2(R) = \frac{1}{2\pi^2}\int_0^{\infty}dk ~k^2 
~W^2_{TH}(kR) ~P(k)\lb{sigW}~.
\ee
Here, $P(k)\equiv\langle\vert\delta_k\vert^2\rangle$ 
is the primordial power spectrum. The averages $\langle \ldots
\rangle$ refer in principle to quantum expectation values, 
but an effective quantum-to-classical transition takes
place \cite{qtoc} so that it is sufficient for our purpose
to deal with effectively classical averages. 

Usually what is meant by the primordial power spectrum is the
power spectrum on superhorizon scales. On these scales, the 
scale dependence of the power spectrum is unaffected by cosmic 
evolution. On subhorizon scales, however, this is not the case, 
and its deformation is correctly described by the adequate 
transfer function (see Eq.~(\ref{T}) below). Hence, the power spectrum $P(k)$ 
on sub-horizon scales
appearing in (\ref{sigW}) must involve convolution with the
transfer function.

The function 
$W_{TH}(kR)$ stands for the Fourier transform 
of the top-hat window function divided by the 
probed volume $V_W=\frac{4}{3}\pi R^3$,
\be
W_{TH}(kR)=\frac{3}{(kR)^3}\bigl (\sin kR-kR\cos kR\bigr )\ .
\ee
The top-hat window function is the most physical choice
to study the formation of PBHs. The reason is that with this choice
the smoothed density contrast $\delta$ describes directly the
average density contrast in the region relevant for PBH formation,
i.e. a sphere of given radius.
The use of a Gaussian window function, for example, would
erroneously yield too small values for $\sigma^2(R)$, see for example
Fig.~4.1 in the textbook by Liddle and Lyth \cite{LL}.

It then follows that the probability $\beta(M_H)$ that a region
 of comoving size $R=(H^{-1}/a)|_{t_k}$ has an averaged 
density contrast at horizon crossing $t_k$ in the range 
$\delta_{min}\leq\delta\leq\delta_{max}$, is given by
\be
\label{beta}
  \beta(M_H)=\frac{1}{\sqrt{2\pi}\,\sigma_H(t_k)}\,
           \int_{\delta_{min}}^{\delta_{max}}\,
           e^{-\frac{\delta^2}{2 \sigma_H^2(t_k)}}\,\textrm{d}\delta
          \approx\frac{\sigma_H(t_k)}{\sqrt{2\pi}\,\delta_{min}}
           e^{-\frac{\delta_{min}^2}{2 \sigma_H^2(t_k)}}\ ,
\ee
where $\sigma_H^2(t_k):=\sigma^2(R)\big|_{t_k}$.

The expression (\ref{beta}) for $\beta(M_H)$ is usually interpreted as
giving the probability that a PBH will be
formed with a mass $M_{PBH}\geq M_H(t_k)$, i.e., greater or
equal to the mass contained inside the Hubble volume when this scale
re-enters the Hubble radius. Strictly speaking, however, 
this is \emph{not} true, since (\ref{beta})
does not take into account those regions that are underdense on a scale $M_H$, 
but nevertheless overdense on some larger scale
(see e.g. the papers by Carr in \cite{CH74} for a discussion).
In the Press-Schechter formalism this seems to be taken care of
in some models by multiplying (\ref{beta}) with a factor $2$. 
Fortunately, in most cases $\beta(M_H)$ is a very rapidly falling function 
of mass, so this effect can be neglected. 
In this case, $\beta(M_H)$ \emph{does} give the probability 
for PBH formation and thus also (at time $t_k$) 
the mass fraction of regions that will evolve
into PBHs of mass greater or equal to $M_H$.

The main problem in calculating the production rate for PBHs is the
correct evaluation of (\ref{sigW}), i.e. to find the direct relation between
the primordial power spectrum and the mass variance. For this we have to
evaluate the power spectrum at the horizon crossing time $t_k$,
giving $P(k,t_k)$. Instead of $P(k)$ itself, one also uses
\be
\delta^2_H(k,t_k)\equiv \frac{k^3}{2\pi^2}~P(k,t_k) = 
          \frac{2}{9 \pi^2}~k^3~\Phi^2(k,t_k)~,\lb{delk}
\ee
where $\Phi$ denotes the gauge-invariant gravitational potential.
We will sometimes use the short-hand 
notation $\delta^2_H(k,t_k)\equiv \delta^2_H(t_k)$. 
Inserting this into (\ref{sigW}), one finds the following expression,
where all quantities are evaluated {\em at the time} $t_k$
\cite{BKP1,P02},
\bea
\sigma^2_H(t_k) &\equiv& \alpha^2(k)~\delta^2_H(t_k)~,\lb{alpha1}\\
&\equiv& \frac{2}{9\pi^2}~\alpha^2(k)~k^3\Phi^2(k,t_k) \lb{alpha2}
\eea
with
\be
\alpha^2(k)= \delta^{-2}_H(t_k)~\int_0^{\frac{k_e}{k}} x^{3}~\delta_H^2(t_{kx})
 T^2(kx,t_k)~W^2_{TH}(x)~\textrm{d}x~.\lb{alphag}
\ee
The upper integration limit is given by the Hubble radius at the end
of inflation, corresponding to the smallest scale generated by inflation.
The transfer function $T(k,t)$ is defined through
\be\lb{T}
P(k,t)= \frac{P(0,t)}{P(0,t_i)}~P(k,t_i)~T^2(k,t)\; , \quad T(k\to 0,t)\to 1~.
\ee
Here, $t_i$ is some initial time when all scales are outside the Hubble radius, $k<aH$, 
and we can take $t_{i}=t_e$ (the time of the end of inflation).

\section{PBHs from primordial spectra}

\subsection{General results with scale-free primordial spectra}

In this paper, we will derive accurate results for the 
production of PBHs from primordial fluctuations 
spectra possessing a characteristic scale. 
We are thus interested in computing the variance $\sigma_H^2(t_k)$ 
at horizon crossing using the 
power spectrum of the primordial fluctuations $P(k)$ of interest to us.
In order to review the formalism and the results already obtained earlier, 
it is convenient to consider first 
the case of a primordial power spectrum $P(k)$ which is scale-free 
and of the form $P(k)=A(t)k^n$ 
on scales larger than the Hubble radius (``superhorizon'' scales). 
Then, from (\ref{alpha1},\ref{alpha2}), we can relate $\sigma_H(t_k)$ to 
the present value $\delta_H(t_0)$ \cite{BKP1},
\bea 
\sigma^2_H(t_k) &=& \frac{100}{81}\alpha^2(k)~\delta_H^2(t_0)~
   \Biggl \lbrack \frac{M_H(t_0)}{M_H(t_{eq})} \Biggr \rbrack^{\frac{n-1}{3}} 
   \Biggl \lbrack  \frac{M_H(t_{eq})}{M_H(t_k)}
   \Biggr \rbrack^{\frac{n-1}{2}}\lb{del}\\
&=& \frac{200}{9^3\pi^2}\alpha^2(k)~k_0^3~\Phi^2(k_0,t_0)~
   \Biggl \lbrack \frac{M_H(t_0)}{M_H(t_{eq})} \Biggr \rbrack^{\frac{n-1}{3}} 
   \Biggl \lbrack  \frac{M_H(t_{eq})}{M_H(t_k)}  \Biggr \rbrack^{\frac{n-1}{2}}~,\lb{sig}
\eea
where $M_H(t_k)$ denotes the horizon mass at horizon entry of the scale
$k$, and $M_H(t_{eq})$ the horizon mass at matter-radiation equality.
The quantity $k_0$ ($t_0$) corresponds to the present Hubble radius
(the present time).  
We have tacitly assumed in (\ref{del},\ref{sig}) that the PBH forms 
before the time of equality, 
$t_k\ll t_{eq}$, and that the universe is first radiation-dominated and 
instantaneously becomes matter-dominated (with $\Lambda=0$). 
A more complicated evolution of the scale factor $a(t)$ 
will result in different expressions for (\ref{del},\ref{sig}). 
Note that the relationship $M_H(t_k)\propto k^{-2}$ for $t_k\ll t_{eq}$ can 
be made slightly more accurate than that used in (\ref{del},\ref{sig}) \cite{P02}. 

 From (\ref{alphag}) we now find that $\alpha$ is given for a
scale-free spectrum by the expression
\be
\alpha^2(k)= \int_0^{\frac{k_e}{k}} x^{n+2}~
 T^2(kx,t_k)~W^2_{TH}(x)~\textrm{d}x~.\lb{alpha}
\ee
The quantities $\delta_H(t_k)$ and $k^{\frac{3}{2}}\Phi(k,t_k)$ scale like 
$\propto k^{\frac{n-1}{2}}$, up to the coefficient 
($\frac{10}{9}$ in our case) 
which is due to the transition between radiation and matter domination. 
However, due to the non-trivial integral (\ref{alpha}),
this is {\it not} the case for the quantity $\sigma_H(t_k)$.

We see from (\ref{alpha1},\ref{alpha2}) that the function 
$\alpha(k)$ introduces an 
additional dependence on $k$. Therefore, it will have two important effects: 
First, 
it affects the value of $\sigma^2_H(t_k)$ when extrapolated from the 
present values 
$\delta_H(t_0)$, or $k_0^3~\Phi^2(k_0,t_0)$, back to the formation time.
But in addition, it will affect the shape of
$\sigma^2_H(t_k)$ as a function of $M_H(t_k)$. 
For example, as we will see in detail, 
a pure step in the primordial power spectrum does not translate anymore into a pure step 
in the quantity $\sigma^2_H(t_k)$. This second aspect is particularly important when one 
considers primordial spectra with a characteristic scale, as we do
in the present paper. 

The observational input needed in (\ref{del}) or (\ref{sig}) is the numerical value of 
the quantity $\delta_H(t_0)$, or $k_0^3~\Phi^2(k_0,t_0)$, on the present Hubble radius 
scale which is found using the CMB 
anisotropy data for large angular scales. 
 Once $k_0^3~\Phi^2(k_0,t_0)$ or equivalently $\delta^2_H(k_0,t_0)$ is 
a known number, the overall normalization of the spectrum is fixed. 
We will take \cite{BW97}
\be
k_0^3~\Phi^2(k_0,t_0)= 1.67 \times 10^{-8}\times {\rm exp}[-0.959(n-1)-0.169(n-1)^2]~,\lb{norm1}
\ee
which corresponds to a flat critical density universe ($\Omega_{m,0}=1$) and we have tacitly 
assumed here that the primordial gravitational waves background is negligible.
%
%
We have in particular
\be
k_0^3~\Phi^2(k_0,t_0)=\frac{9 \pi^2}{2} \delta^2_H(t_0)~. \lb{norm3}
\ee
We should mention here the effect of a cosmological constant, as recent observations strongly 
suggest that we live in a nearly flat universe with 
$\Omega_{\Lambda,0}\approx 0.7$ and $\Omega_{m,0}\approx 0.3$. 
In such a Universe, the effect of including a cosmological constant would merely reduce 
the mass variance $\sigma_H(t_k)$ by about 15\% \cite{P02}.
In a more general quintessence model, this decrease is model-dependent. 

\subsection{Primordial spectrum with a pure step}

We consider now more sophisticated spectra with a 
characteristic scale for which 
Eqs.~(\ref{del},\ref{sig},\ref{alpha}) do not apply anymore. For each specific spectrum 
we have to compute $\sigma_H(t_k)$ numerically, but it is possible 
to give a rough estimate of what will happen for a class of primordial spectra 
using some simple models. In \cite{BKP1} 
we have discussed the simplest case in which
the step occurs directly in the mass variance $\sigma_H(t_k)$. Here we
consider a pure step in the primordial power spectrum  
at $k=k_s$, with the corresponding time of re-entrance
satisfying $t_ {k_s}<t_{eq}$,
\be
k^3~\Phi^2(k,t_k)= \left (\frac{10}{9} \right )^2~k_0^3~\Phi^2(k_0,t_0)~
\left (\frac{k}{k_0} \right )^{n-1}~\times 
\left\{ \begin{array}{ll} 
                  1      & \textrm{for $k<k_s$} \\ 
                  p^{-2} & \textrm{for $k\geq k_s$} 
     \end{array} \right.\, , \ \lb{phip}
\ee
where the ratio of power on large scales 
to that on small scales is given by the parameter $p^2$.
Note that we assume that aside from the step, the spectrum has a 
constant spectral index $n$. 

One finds
\bea 
\sigma_H^2(t_{k})&=& \frac{200}{9^3\pi^2}~{\tilde\alpha}^2(k)~k_0^3\Phi^2(k_0,t_0)
         \Biggl \lbrack \frac{M_H(t_0)}{M_H(t_{eq})} \Biggr \rbrack^{\frac{n-1}{3}} 
         \Biggl \lbrack  \frac{M_H(t_{eq})}{M_H(t_k)} \Biggr \rbrack^{\frac{n-1}{2}} \lb{phipg1} \\ 
&=& \frac{100}{81}~{\tilde\alpha}^2(k)~\delta^2_H(t_0) 
         \Biggl \lbrack \frac{M_H(t_0)}{M_H(t_{eq})} \Biggr \rbrack^{\frac{n-1}{3}} 
         \Biggl \lbrack  \frac{M_H(t_{eq})}{M_H(t_k)} \Biggr \rbrack^{\frac{n-1}{2}}~,\lb{delpg1}
\eea
where the quantity $\tilde\alpha(k)$, which is slightly different from the quantity $\alpha(k)$ 
defined in (\ref{alpha1}), is given by 
\bea
{\tilde\alpha}^2(k)&=&\int_0^{\frac{k_s}{k}} x^{n+2}~
 T^2(kx,t_k)~W^2_{TH}(x)~\textrm{d}x\\ \nonumber
&+&p^{-2}\int_{\frac{k_s}{k}}^{\frac{k_e}{k}} x^{n+2}~
 T^2(kx,t_k)~W^2_{TH}(x)~\textrm{d}x~.\lb{alphaptilde}
\eea
It is seen that the location of the step introduces a dependence on $k$.
It is also seen that, as must be the case, the effect of the step  
disappears both for $k_s\to k_e$ and $k_s\to 0$. The difference between both limits 
corresponds just to a different overall normalization.

\subsection{BSI spectrum with large oscillations}

We now further generalize the spectra under consideration and consider 
a primordial spectrum produced in an 
inflationary model with a jump in the first derivative of the 
inflaton potential 
$V(\phi)$ at some scale $k_s$ and corresponding value $\phi_s$
of the inflaton. 
The resulting spectrum for such a potential is of a universal form, 
and an exact analytical expression has been derived by Starobinsky \cite{S92}.
At re-entrance 
inside the Hubble radius during the radiation dominated stage, one has the following 
primordial spectrum,
\be \label{Phi}
k^3 \Phi^2(k,t_k)\equiv \frac{4}{9} 
~F(k)\ ,~~~~~~~~~~~~~~~~t_{k_e}\ll t_k\ll t_{eq}\ ,
\ee
where 
\bea
F(k)=\frac{9H^6_s}{2A_-^2}~\Biggl [1-3(p-1)\frac{1}{y}\left( \left(1-\frac{1}{y^2} \right) \sin2y + 
\frac{2}{y}\cos2y\right) \nonumber\\
+\frac{9}{2}(p-1)^2\frac{1}{y^2} \left(1+\frac{1}{y^2}\right)
\left(1+\frac{1}{y^2}+\left(1-\frac{1}{y^2}\right) \cos2y
-\frac{2}{y}\sin 2y\right)\Biggr ] \lb{F}
\eea
with
\be
y=\frac{k}{k_s}, \qquad p=\frac{A_-}{A_+}, 
\qquad H_s^2=\frac{8\pi G V(\phi_s)}{3}\ ,
\nonumber
\ee
where $\Phi$\ is as above 
the (peculiar) gravitational potential and the quantities $A_-,~A_+$ are 
the inflaton potential derivatives on both sides of the jump. 
This expression depends (besides the overall normalization) on two parameters $p$\ and $k_s$.
The shape of the spectrum depends only on $k/k_s$; the scale $k_s$
only determines the location of the step. Note that we have 
\bea
F(0) &=& \frac{9H_s^6}{2A_+^2}\ ,\\
F(\infty) &=& \frac{9H_s^6}{2A_-^2}\equiv \frac{F(0)}{p^2}\ .\lb{F0}
\eea
For $p>1$, the spectrum has a flat upper plateau on larger scales, 
even with a small bump, and a sharp decrease on smaller scales, with large 
oscillations. Hence, compared to a pure step, the spectrum (\ref{F}) is a step which 
is dressed up with a rich structure, in particular large oscillations confined to 
the neighborhood of the step.   
But one can also have the case with $p<1$ with an increase on small scales. 
This will be the case of interest to us, and this possibility with the corresponding
variance is displayed in Fig.~1.

\begin{figure}[t]
  \begin{center} 
\mbox{\input{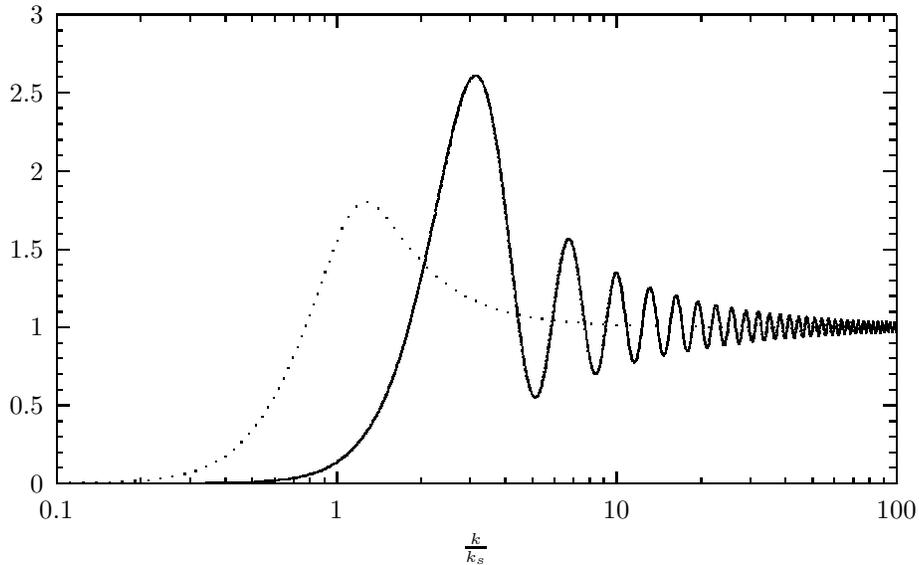}}
  \end{center}
  \caption[]{ The mass variance $\sigma_H^2(t_{k})$ (dotted line) and the primordial power 
spectrum $k^3\Phi^2(k,t_k)\propto \delta^2_H(t_k)$ (solid line) for the Starobinsky-type 
BSI spectrum (\ref{F}) are displayed for $p=7.58\times 10^{-4}$. The normalization of each 
curve is arbitrary. This particular value of the parameter $p$ corresponds to $p_{\gamma}$, 
the absolute minimal value allowed by the constraint set by the contribution of evaporated 
PBHs to the diffuse $\gamma$-ray background (see below).}
  \label{bsispectrum}
\end{figure}

The power spectrum at re-entrance inside the Hubble radius is then given by 
\begin{equation}
k^3 P(k,t_k)=\left (\frac{2}{3}\right )^4 F(k) ~.
\end{equation}
We have in particular
\be
k_0^3~\Phi^2(k_0,t_0)= \frac{9 \pi^2}{2} \delta^2_H(t_0)=\frac{9}{25}~F(k_0)
                     \simeq \frac{9}{25}~F(0)~\lb{norm2}
\ee
for big scales.
Using (\ref{F0},\ref{norm2}), we can write the normalized spectrum explicitly.
For this purpose it is convenient to introduce the notation 
\be
F(k)\equiv \frac{9H^6_s}{2A_-^2}~f(k)=\frac{F(0)}{p^2}~f(k)~.\lb{fk}
\ee

One is then finally led to the following result,
\be
\sigma_H^2(t_{k})=
\frac{100}{81}p^{-2}~\delta_H^2(t_0)~\int_0^{\frac{k_e}{k}} 
f(kx)~x^{3}~T^2(kx,t_k)~
                         W^2_{TH}(x)~\textrm{d}x~\lb{sigFd}.
\ee
Though we have introduced here the function $F(k)$ in relation to the particular BSI 
spectrum considered here, it actually applies to spectra of arbitrary shape, whether
scale-free or not, with $F(k)$ defined through (\ref{Phi}). Indeed, we 
then obtain 
\be
\sigma_H^2(t_{k})=\frac{8}{81\pi^2} \int_0^{\frac{k_e}{k}} x^{3}~F(kx)~T^2(kx,t_k)~
                                            W^2_{TH}(x)~\textrm{d}x~,\lb{sigF} 
\ee
cf. our general formula (\ref{alpha2}).
It is easily checked that the general expression (\ref{sigF}) reduces to the expressions 
derived earlier when $F(k)=Ck^{n-1}$ for a scale-free powerlaw spectrum or when it has a superimposed pure step. 
Finally, in order to study PBH formation, it is physically appealing to express the results of this 
section in terms of the mass $M_H(t_k)$ instead of the wavenumber $k$, as is actually done 
in (\ref{del},\ref{sig}). 
We have in particular (for $t_k,t_{k_s}< t_{eq}$)
\be
  \frac{k}{k_s}=\sqrt{\frac{M_H(t_{k_s})}{M_H(t_k)}}~.
  \lb{kM2}
\ee
For all the primordial spectra considered in this Section, we will derive now accurate expressions 
for $\sigma_H^2(t_{k})$ and then evaluate them numerically.
 
\subsection{Accurate expression for $\sigma_H^2(t_{k})$}

We finally have to deal with the transfer function $T(k',t_k)$ in order 
to have an accurate evaluation of the integral expression for the mass variance 
$\sigma_H^2(t_{k})$. 
In the radiation era, $\delta_k$ behaves like 
\be
\delta_k\propto \frac{\sin (\mu a)}{\mu a} - \cos (\mu a)\lb{deltak}\ ,
\ee
where the constant $\mu$ is given by  
\be
\mu\equiv \frac{k c_s}{a^2 H}~~~~~~~~~~~~~~~~c_s=\frac{1}{\sqrt{3}}~.
\ee
It is easily checked that (\ref{deltak}) has the right behavior on scales larger than 
the Hubble radius,
\be
\delta_k \propto a^2~~~~~~~~~~~~~~~~k\ll aH~. 
\ee
Therefore, after matching on large scales, 
the following expression is obtained, 
\be\lb{sigF1}
\sigma_H^2(t_{k})=\frac{8}{81\pi^2}\int_0^{\frac{k_e}{k}} x^3~F(kx)~
\Biggl[ \frac{9}{x^2}\left(\frac{\sin(c_s x)}{c_s x}-\cos(c_s x)\right) \Biggr]^2~
                                        W^2_{TH}(x)~\textrm{d}x~, 
\ee
which is valid for {\it any} primordial spectrum, with $F(k)$ defined in (\ref{Phi}) 
independently of the particular expression (\ref{F}) for $F(k)$ that we consider here. 
For a scale-free powerlaw spectrum, $F(k)=Ck^{n-1}$, expression (\ref{sigF}) reduces to 
\be\lb{sign}
\sigma_H^2(t_{k})= \delta_H^2(t_k) \int_0^{\frac{k_e}{k}} x^{n-2}~
\Biggl[ 9\left(\frac{\sin(c_s x)}{c_s x}-\cos(c_s x)\right) \Biggr]^2~W^2_{TH}(x)~\textrm{d}x~.
\ee
By inspection of (\ref{sigF},\ref{sigF1}), and 
(\ref{alpha1},\ref{alpha},\ref{sign}), we have the following identification,
\be
T^2(kx,t_k)\equiv \Biggl[\frac{9}{x^2}\left(\frac{\sin(c_s x)}{c_s x}-\cos(c_s x)\right) \Biggr]^2~=
W^2_{TH}(c_sx)= W^2_{TH}\left(\frac{x}{\sqrt{3}}\right)~,
\ee 
so that (\ref{sigF1}) and (\ref{sign}) can be written as 
\be\lb{sigFW}
\sigma_H^2(t_{k})=\frac{8}{81\pi^2} \int_0^{\frac{k_e}{k}} x^3~F(kx)~W^2_{TH}(c_sx)~W^2_{TH}(x)~
                                                   \textrm{d}x~,
\ee
and
\be\lb{signW}
\sigma_H^2(t_{k})= \delta_H^2(t_k) \int_0^{\frac{k_e}{k}} x^{n+2}~
                                  W^2_{TH}(c_sx)~W^2_{TH}(x)~\textrm{d}x~,
\ee
respectively.
For a scale-free powerlaw primordial spectrum, 
the following numerical value for $\alpha^2(k)$ is obtained:  
\be\lb{alphanum}
5.37\leq \alpha^2(k) \le 6.83\ ,~~~~~~~~~~~~~~~~~~~~~1\leq n \leq 1.3\ .
\ee
So, even for a pure Harrison-Zel'dovich ($n$=1) spectrum, 
there is a decrease on small scales compared to the
normalization used earlier in the literature, with now
\be
\sigma^2_H(t_k)= 6.63 ~\delta^2_H(t_0)\ ,~~~~~~~~~~~~~~~n=1~,
\ee
because the scaling $\sigma^2_H(t_k)\propto k^{n-1}$ would be inaccurate.
The normalization used earlier in the literature corresponds to 
$\sigma^2_H(t_k)\approx 25 ~\delta^2_H(t_0)$ \cite{GL}.

Let us remark that in \cite{BKP1} we 
have taken as a rough estimate $T^2(x<1)=1,~T^2(x\geq 1)=x^{-4}$. We see that this is excellent for 
$0\leq x \leq 1$; however, the amplitude of the oscillations inside the 
Hubble radius is larger than 
its value at Hubble radius crossing and thus requires the accurate 
analytic calculation done here. 

\begin{figure}[t]
  \begin{center} 
     \psfrag{k}[][][0.8]{$k/k_s$}
     \psfrag{s}[][][0.8]{$\sigma^2_H(t_k)$}
     \psfrag{l1}[][][0.8]{BSI}
     \psfrag{l2}[][][0.8]{pure step}
  \includegraphics[width=0.7\textwidth]{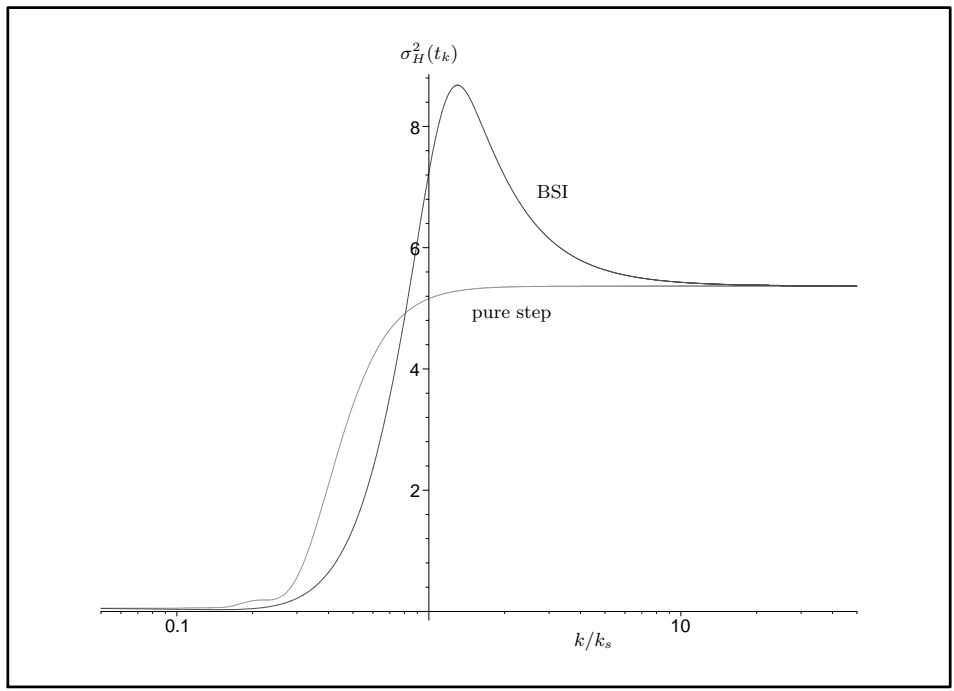} 
  \end{center}
  \caption[]{This picture shows the mass variance $\sigma_H(t_k)$ for 
two scale-dependent 
primordial spectra with the same ratio of power between small and large scales, i.e. the same $p$: 
for the pure step, Eqs.~(\ref{phipg1},\ref{delpg1}), and for the BSI spectrum,
Eqs.~(\ref{sigFd}), both with $n=1$ and $p=0.1$.} 
\label{sigmapic}
\end{figure}

We see that actually $T^2(kx,t_k)$ does not depend on $k$. This will be true as long as $\delta_k$ 
on all scales relevant in the integration is given by (\ref{deltak}). This is the case for 
scales of interest here corresponding to PBH formation very deep inside the radiation era 
and $M_{PBH}\geq M_*$, and it would remain true for much lower masses.  
Therefore, for a powerlaw scale-free spectrum, the only dependence on $k$ comes from the upper limit 
of integration. 
However, as the integrand decreases rapidly, for all PBHs with 
$M\geq M_*$, this limit can be safely replaced by infinity. 
Hence, we conclude 
that for a scale-free powerlaw primordial spectrum, the function $\alpha$ is essentially 
independent of $k$. 
Incidentally, whenever $\alpha$ does not depend on $k$, our improved normalization is formally 
equivalent to taking the ``old'' normalization and a new significantly larger $\delta_{min}$. 
In order words, there is a degeneracy between taking a smaller constant $\alpha$ with fixed 
$\delta_{min}$ and taking a larger $\delta_{min}$ with fixed constant $\alpha$, provided they satisfy 
\be
\left( \frac{\alpha}{\delta_{min}}\right )_{old} = \left( \frac{\alpha}{\delta_{min}}\right )_{new}~.
\ee
Still, a dependence on $k$ can appear due to a feature in the primordial 
spectrum, as can be inferred from (\ref{sigF}). But the interesting point is that due to 
the presence of $\alpha(k)$, the scale dependence of the primordial spectrum is translated 
in a non-trivial way into the scale dependence of the mass variance at horizon crossing.
Fig.~\ref{sigmapic} shows the mass variance for the BSI primordial spectrum 
(\ref{sigFd}), and for the primordial spectrum with a pure step 
(\ref{phipg1},\ref{delpg1}), both with $n=1$. 
It is seen that the pure step in the primordial spectrum gets smoothed 
when it comes to the 
mass variance. Hence, even if one assumes that the PBH mass $M_{PBH}$ is exactly the Hubble 
mass $M_H(t_k)$, a pure step at the scale $k=k_s$ will boost the production of PBH on a certain 
range of masses $\Delta M$ in the neighborhood of $M_s\equiv M_H(t_{k_s})$. In the example shown 
on Fig.~\ref{sigmapic}, the rise in the mass variance starts already around 0.05 $M_s$ and 
the asymptotic value is reached around 5 $M_s$.

As for the BSI primordial spectrum (\ref{sigFd}), we see that the rich structure of the
spectrum (\ref{F}) has nearly completely disappeared due to the effects of the filtering. 
Nevertheless, and most importantly, what remains is a noticeable peak in $\sigma_H(t_k)$. 
Note that the peak is not found at $k_s$, but at a slightly smaller 
scale with $k_{peak}\approx 1.27~k_s$.
In the past, this BSI spectrum was used in order to explain the presence of a bump in the matter 
spectrum on a scale around 125$h^{-1}$Mpc \cite{LPS98}.
Here, we have a completely different possibility: such a spectrum could produce a fairly
localized bump in the mass variance at much lower scales.    

We will now put some constraints on the parameter values of our spectra.

\section{Observational constraints}

There are various limits on the initial mass fraction
 $\beta(M_H)$ of PBHs of mass $M_{PBH}\geq M_H$, most of them being related to 
Hawking radiation. Here, we consider first the gravitational constraint 
\cite{BKP1}
\begin{equation}
  \label{gc2}
  \beta(M_H) < 1.57 \times 10^{-17}\left(\frac{M_H}{10^{15}\textrm{g}}\right)^
\frac{1}{2}~h^2,
\end{equation}
which comes from the requirement that the PBH mass density today must not exceed the 
present (critical) density of the universe, $\Omega_{PBH,0}<1$. 
Such a constraint was derived, for example, in \cite{CGL,GL}.
Note, however, that the numerical factor in (\ref{gc2}) 
makes use of the adiabatic expansion of the universe and all the relevant 
$g$ factors \cite{BKP1}. We emphasize also that in the derivation of (\ref{gc2}), the quantity 
$\beta(M_H)$ is interpreted as being the fraction of the energy density that forms PBH of mass $M_H$.  
Actually, we will be interested in several possible values of the quantity $\Omega_{PBH,0}$.
The generalized version of (\ref{gc2}) can be written in the following way 
\be
\Omega_{PBH,0}(M)~h^2 = 6.35 \times 10^{16}~\beta(M) 
           ~\left(\frac{10^{15}\textrm{g}}{M}\right)^\frac{1}{2}~.\lb{gc3}
\ee  
Since PBHs of initial mass $M_{PBH}\lesssim M_*$ will have 
evaporated by the present day, the tightest constraint is 
obtained for the PBH with smallest mass not yet evaporated, namely $M_H\sim 10^{15}$ g.

In this section, we will derive constraints on the parameters of the power spectra introduced above. 
In order to do this the quantity $\sigma_H(t_k)$ has to be evaluated numerically, 
then inserted in (\ref{beta}) in order to find $\beta(M_H)$, and finally constrained using 
(\ref{gc2},\ref{gc3}). 

In the following we shall use
\begin{equation}
  \label{par}
  h=0.7\ ,~~~~~~~\qquad\delta_{min}=0.7,
\end{equation}
and evaluate (\ref{beta}) at $M_H=10^{15}$ g. 

\subsection{Scale-free spectrum}

The case of a scale-free spectrum was already considered in \cite{BKP1} 
using an estimate for $\alpha(k)$, namely $\alpha(k)\approx 1.5$ for the relevant masses 
$M_{PBH}\sim 10^{15}$ g. The following upper bound on the spectral index $n$ is then 
found for the parameters (\ref{par}) (Fig.~1 in \cite{BKP1} corresponds to 
$h=0.5,~\delta_{min}= 1/3$):
\begin{equation}
  \label{n1}
  n\lesssim 1.36~.
\end{equation}
Performing now a more accurate calculation based on (\ref{signW}), corresponding to 
\be
\alpha^2(k) = \int_0^{\frac{k_e}{k}} x^{n+2}~
                        W^2_{TH}(c_sx)~W^2_{TH}(x)~\textrm{d}x~,
\ee
and for the values given in (\ref{par}), we refine the upper bound to the value
\begin{equation}
  \label{n2}
  n\lesssim 1.33~.
\end{equation}
We have discussed in \cite{BKP1} the dependence of this result 
on the parameters given above: the dependence on $h$ is very weak, while 
that on $\delta_{min}$ is significant.
We should stress that the constraint derived here on scale-free spectra 
is rather academic in the sense that the production of PBHs with mass smaller than 
$M_*$, whose evaporation would contribute to the $\gamma$-ray background, 
would be higher by many orders of magnitude than that allowed. Of course, the upper 
bound we have found using the gravitational constraint alone could be relevant if the 
mass scale of inflation corresponds to $M_*$.

\subsection{Spectrum with a pure step}

\begin{figure}[t]
  \begin{center} 
     \psfrag{m}[][][0.8]{$\lg M$}
     \psfrag{b}[][][0.8]{$\lg \beta(M)$}
  \includegraphics[width=0.7\textwidth]{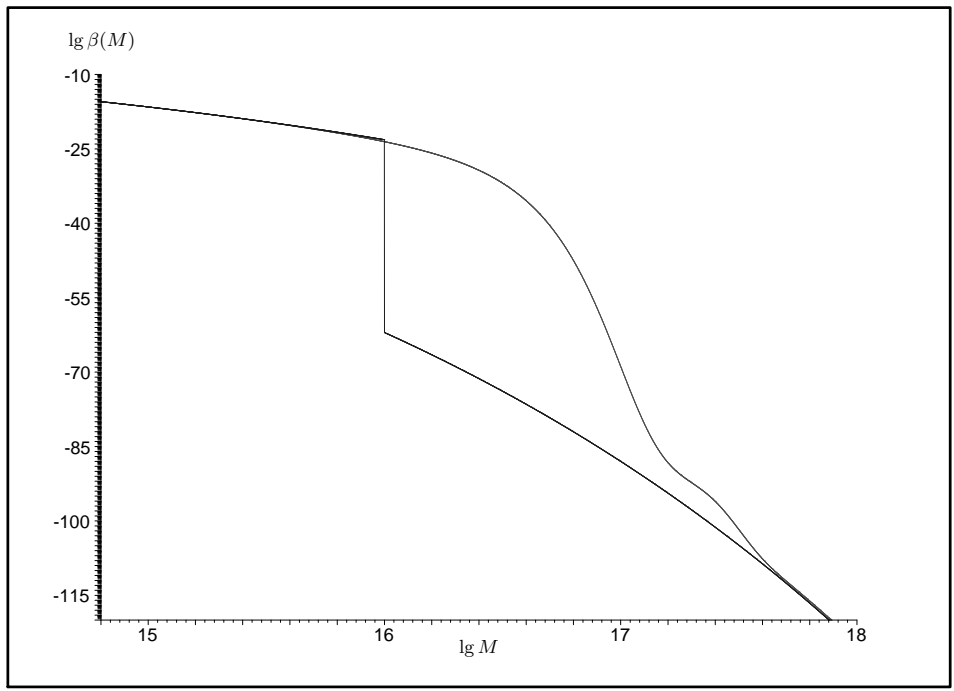} 
  \end{center}
  \caption[]{The quantity $\beta(M)$ is shown for the spectrum (\ref{phipg1},\ref{delpg1}),
i.e. a pure step in the primordial power spectrum itself, and for the case of a pure 
step in $\sigma_H(t_k)$ (as discussed in \cite{BKP1}). Both correspond to (arbitrarily chosen)
parameters $n=1.31$, $p=0.6$ and $M_s\equiv M_H(t_{k_s})=10^{16}$ g. Both spectra shown here satisfy
the gravitational constraint, Eq.~(\ref{gc2}), which applies only for $M\gtrsim10^{15}$ g.}\label{stepbeta}
\end{figure}

In the following, we shall restrict ourselves to the case $p\leq 1$, i.e., more power on 
small scales. This is the physically interesting case for us, since it offers the 
possibility to produce more instead of less PBHs on small scales, which can have 
interesting cosmological constraints. Furthermore, the opposite case leads to 
interpretational problems for the quantity $\beta(M)$: For $p>1$ it is no longer a 
monotonically falling function of mass and therefore 
cannot be interpreted anymore as the fractional mass density of PBHs of mass 
larger than $M$. For such a case, one must be cautious with the interpretation of the 
Press-Schechter formalism.

Fig.~\ref{stepbeta} shows explicitly the important consequence of the non-trivial relationship 
between the quantities $\sigma_H(t_k)$ and $\delta_H(t_k)$ mentioned above. Even though the 
primordial spectrum has a pure step at some scale $k=k_s$, the resulting quantity 
$\sigma_H(t_k)$ does not. Indeed, the convolution of $\delta_H(t_k)$ with the scale dependent 
function $\alpha(k)$ leads to a ``smoothed-out'' step. As a result, 
this ``smoothing'' will also apply to $\beta(M)$, the variable depicted
in Fig.~\ref{stepbeta}.
 This is important whenever one is willing to overshoot 
the PBH production at some given scale: It places an avoidable limit to the sharpness of the ``step'', 
and clearly of any feature, that will survive in $\sigma_H(t_k)$. 
Still, as can be seen from Fig.~\ref{stepbeta}, the 
``step'' remains fairly well localized in the neighborhood of $M_s$ (corresponding to 
$10^{16}$ g in the case displayed on Fig.~\ref{stepbeta}).

\begin{figure}[t]
  \begin{center} 
     \psfrag{p}[][][0.8]{$\lg p$}
     \psfrag{n}[][][0.8]{$n_{max}$}
     \includegraphics[width=0.7\textwidth]{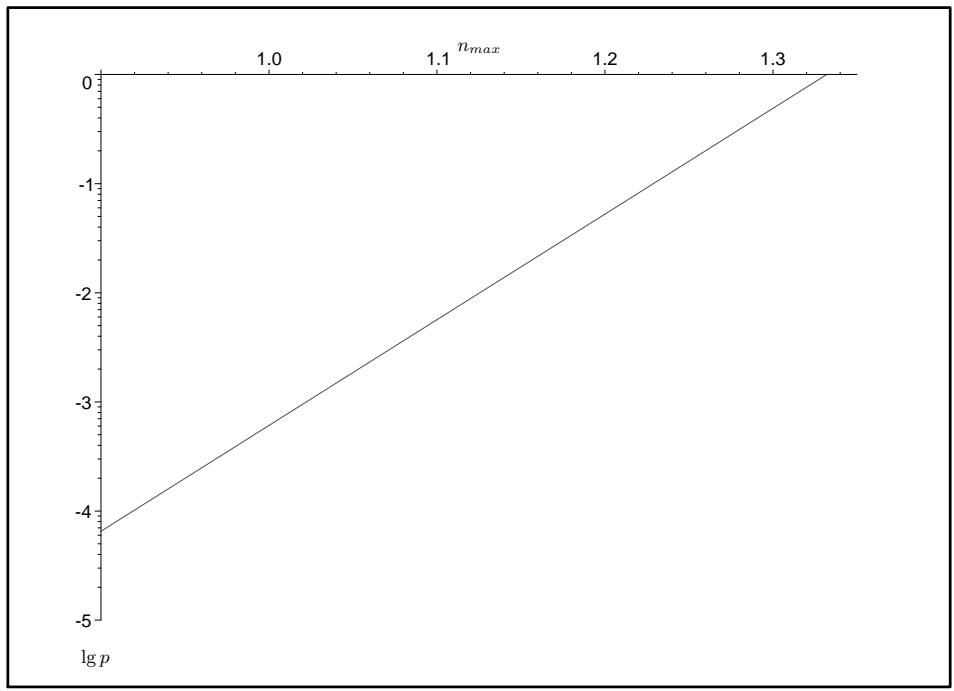} 
  \end{center}
  \caption[]{Dependence of $n_{max}$ on $p$ for a spectrum with a pure step in $\sigma_H(t_k)$
or in $\delta_H(t_k)$ at $M_s\equiv M_H(t_{k_s})\gtrsim10^{15}$ g. For $p=1$ the result for
the scale-free case is recovered, whereas for $p<1$ the constraint on $n$ is considerably strengthened.} 
\label{pn}
\end{figure}

A pure step in $\sigma_H(t_k)$ itself has been considered before 
\cite{BKP1} and leads to a relationship 
between $p$ and the maximal allowed value for the spectral index $n_{max}$ as depicted 
in Fig.~\ref{pn}. Note that the result is independent of the scale $k_s$ (as long as 
$M_s>10^{15}$ g) and that for $p=1$ the result of the scale-free case, 
$n_{max}\approx1.33$, is recovered. No information about the allowed height of the step 
can be obtained if $M_s\lesssim 10^{15}$ g from the condition (\ref{gc2}) for a primordial 
spectrum for which the quantity $\sigma_H(t_k)$ has a step structure, 
since by definition the latter constraint applies only for $M_s\gtrsim 10^{15}$ g, 
which is the condition that non-evaporated PBHs are affected by the step.
As becomes clear from Fig.~\ref{stepbeta}, for $M_s\gtrsim10^{15}$ g one obtains 
the same results for both spectra, i. e. the relationship between $n_{max}$ and $p$ as 
depicted in Fig.~\ref{pn} is independent of whether one has a pure step in $\sigma_H$ or 
in $\delta_H$ itself. In the latter case, however, one also finds a (weak) constraint on 
the allowed height of the step for 
$5\times10^{13}\ \mathrm{ g}\lesssim M_s\lesssim 10^{15}\ \mathrm{ g}$. 
This is due to the ``smoothing out'' of the step.

\subsection{BSI spectrum}

\begin{figure}
  \begin{center} 
\mbox{ \input{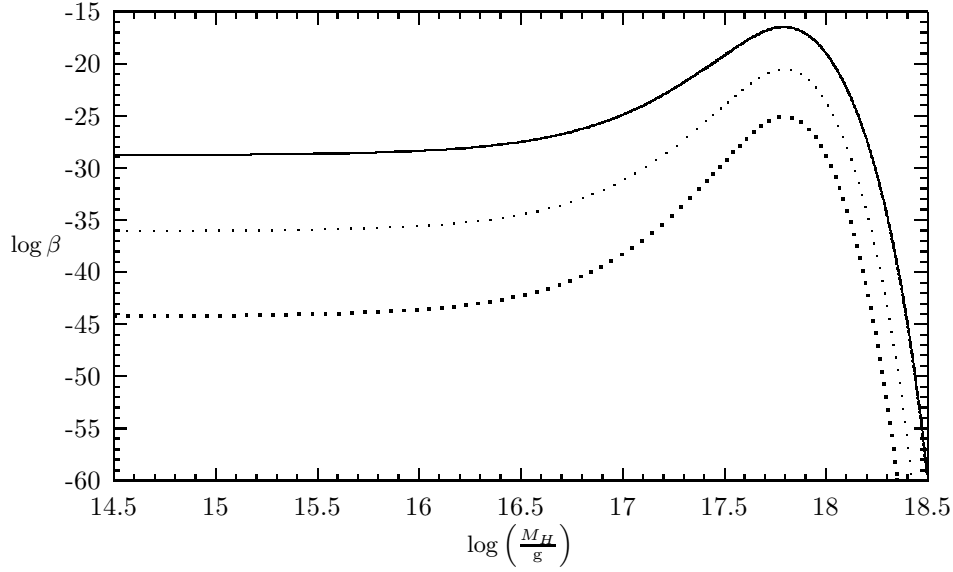} }
  \end{center}
  \caption[]{The quantity $\beta(M_{H})$ 
is shown for the BSI-spectrum (\ref{sigFd})
containing a jump in the inflaton potential derivative for 
the parameters $M_{H}(t_{k_{s}})=10^{18}$g, and from bottom to top
$p=10^{-3},~9\times 10^{-4},~8\times 10^{-4}$. 
As can be seen, $\beta(M_{H})$ acquires a well localized bump 
in the vicinity of $M_s\equiv M_{H}(t_{k_s})$. Note that we have 
$\beta_-\equiv \beta(M_{H}\ll M_{peak})\lesssim 10^{-12}\times \beta(M_{peak})$,
an inequality sufficient to avoid the severe constraint from the contribution to the 
$\gamma$-ray background of evaporated PBHs with $M_{H}\leq M_*$.}
\label{bsibeta}
\end{figure}

Computing $\beta(M_H)$ for the BSI spectrum  (\ref{sigFd}), one discovers that 
it is no longer a decreasing function of mass, but that it has a peak at $M_{peak}$ instead. 
This is illustrated in Fig.~\ref{bsibeta} and constitutes the most interesting result  
of our BSI spectrum in the context of PBH formation. As can be seen from the comparison 
with a primordial spectrum possessing a pure step structure shown in 
Fig.~\ref{sigmapic}, 
the presence of a bump in $\sigma_H(t_k)$ is a consequence of the oscillatory behavior 
of the BSI spectrum in the vicinity of $k_s$. 
Let us note that there is a one-to-one
correspondence between $k_{peak}$ and $k_s$ for each $p$, the ratio is approximately 
given by $k_{peak}\approx 1.27~k_s$, corresponding to $M_{peak}\approx 0.62~M_s$, for the
values of $p$ of interest in the following, namely $p\lesssim \mathcal{O}(10^{-3})$.

\begin{figure}
  \begin{center} 
\mbox{ \input{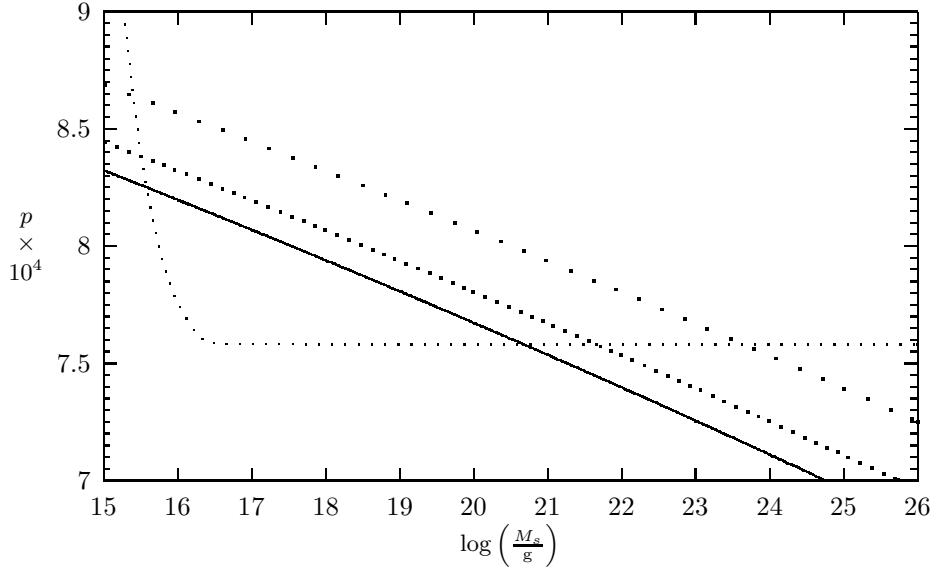} }
  \end{center}
  \caption[]{The allowed region in parameter space $(p, M_s\equiv M_{H}(t_{k_{s}}))$ 
is shown. The solid line represents those points for which $\Omega_{PBH,0}(M_{peak})=0.3$.
Below the solid line we have $\Omega_{PBH,0}(M_{peak})>0.3$. In other words, the gravitational 
constraint at $M_{peak}$ is violated and this region is therefore excluded.
The two lines parallel to the solid line represent, from bottom to top, those points for which 
$\Omega_{PBH,0}(M_{peak})=0.1$ and $0.01$, respectively.
Below the dotted line, the $\gamma$-ray background constraint is violated. It is seen that 
for $M_s\gtrsim 10^{21}$g, the allowed parameter values $p$ 
yield a density $\Omega_{PBH,0}(M_{peak})<0.3$ which becomes rapidly negligible with growing $M_s$.}
\label{constraint}
\end{figure}

A most interesting application of the bump is the possibility to produce a significant 
fraction of the cold dark matter (CDM) in our Universe \cite{BKP2}. 
We shall use the gravitational constraint corresponding to $\Omega_{PBH,0}\le 0.3$. This 
corresponds to a conservative upper bound 
where in our scenario all of CDM is made up of PBHs. 
The condition $\Omega_{PBH,0}\le 0.3$ then implies for $\beta(M_{H})$:
\be
\beta(M_{H})\le 0.48\times 10^{-17} \left(\frac{M_{H}}{10^{15}{\rm g}}\right)^\frac{1}{2}~h^{2}~.
\ee
The strongest gravitational constraint is obviously obtained at $M=M_{peak}$ due to the well 
localized bump. Furthermore, essentially all of the PBHs are produced with $M\approx M_{peak}$.
Hence our gravitational constraint has to be written as follows, 
\be
\Omega_{PBH,0}\approx \Omega_{PBH,0}(M_{peak})\le 0.3~.\lb{gcBSI}
\ee
As can be seen from Fig.~\ref{bsibeta}, for a given $M_s$, taking smaller 
values of $p$ increases $\beta(M_{H})$ and therefore also $\Omega_{PBH,0}$. 
Thus for each $M_s$ there exists a minimum value for $p$, $p_{min}$, in
order to achieve the gravitational constraint $\Omega_{PBH,0}=0.3$.
In this way we obtain a relation between $p_{min}$ and $M_s$ as 
displayed in Fig.~\ref{constraint} by the solid line.
A shift in the location of the characteristic mass $M_s$
(and in the location of the corresponding scale $k_s$) to higher masses $M_s$ 
(equivalently lower $k_s$) still much smaller than
$10^{-7}{\rm M}_{\odot}$ is allowed, but lower values $p_{min}$ are then 
needed in order to achieve $\Omega_{PBH,0}=0.3$.

It is possible to use the observed extragalactic diffuse $\gamma$-ray background to achieve
an interesting window for the characteristic mass scale $M_s$ of our model.
For this we are considering the PBHs which have evaporated by now and may contribute to the
observed $\gamma$-ray background. Their mass lies in the range $\Delta M$,
with $2\times 10^{13}\ \mathrm{ g}\leq M_{H} \leq 5\times 10^{14}\ \mathrm{ g}$
\cite{KL99}.
The $\gamma$-ray background constraint comes from the requirement that the $\gamma$-ray 
flux from PBHs whose mass is in the range $\Delta M$ cannot be larger than the observed diffuse 
$\gamma$-ray background, leading to $\Omega_{PBH,0}(\Delta M)\lesssim 10^{-8}$. 
It is for this reason that the sudden and substantial drop of $\beta(M_{H})$ on 
smaller scales $M_{H}\ll M_{peak}$ is crucial.
From (\ref{gc3}), it is easy to obtain the following equality, 
\be
\frac{\Omega_{PBH,0}(M_{H}\ll M_{peak})}{\Omega_{PBH,0}(M_{peak})}=
\frac{\beta_-}{\beta_{peak}}\times \left(\frac{M_{peak}}{M_{H}}\right)^\frac{1}{2}\lb{eq}
\ee
with $\beta_{-} \equiv \beta(M_{H}\ll M_{peak})$ and $\beta_{peak}\equiv \beta(M_{peak})$.

\begin{figure}
  \begin{center} 
\mbox{ \input{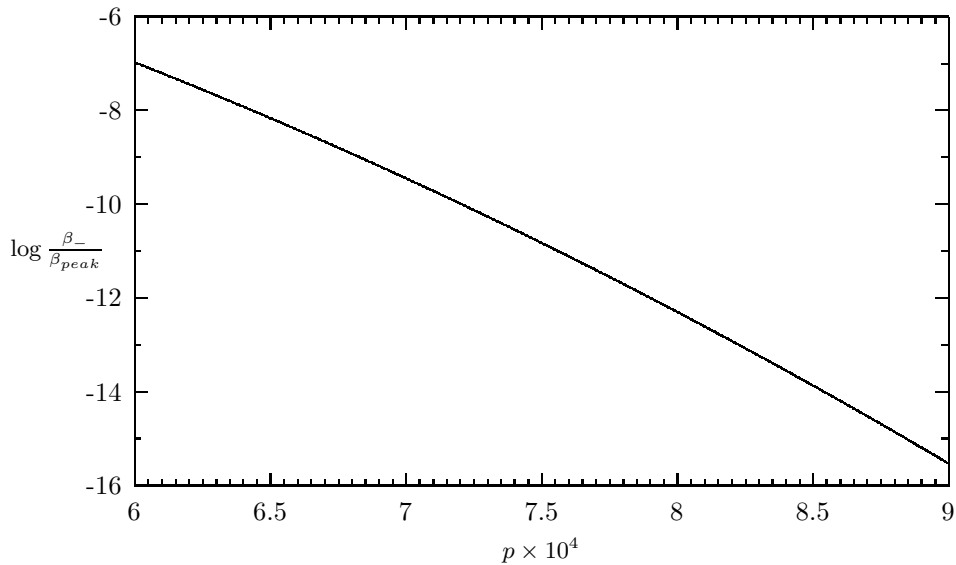} }
  \end{center}
  \caption[]{The ratio $\frac{\beta_-}{\beta_{peak}}$, with $\beta_{peak}\equiv \beta(M_{peak})$, 
is shown as a function of $p$. For $p \gtrsim 7.5\times 10^{-4}$ the ratio is sufficiently
small in order to satisfy the severe upper bound set on the contribution of evaporated PBHs to the
extragalactic diffuse $\gamma$-ray background.}
\label{rapportfig}
\end{figure}

Taking $\Omega_{PBH,0}(M_{peak})=0.3$ 
and using the $\gamma$-ray constraint $\Omega_{PBH,0}(M_{H}\ll M_{peak}) < 10^{-8}$, (\ref{eq}) 
leads to 
\be
\frac{\beta_{-}}{\beta_{peak}} < 3.3 \times 10^{-8}~ \left
 ( \frac{M_{H}}{M_{peak}}\right)^{\frac{1}{2}} \lb{rapport}
\ee
where $M_{H}$ is any mass lying in the range $\Delta M$.\\
Taking smaller values of $p$ increases the ratio $\frac{\beta_{-}}{\beta_{peak}}$, as can be seen 
on Fig.~\ref{rapportfig}. Indeed, using (\ref{beta}) we obtain
\be
\frac{\beta_{-}}{\beta_{peak}}=\frac{\sigma_{-}}{\sigma_{peak}}
\exp\left(-\frac{\delta_{min}^{2}}{2\sigma^{2}_{-}}
\left(1-\frac{\sigma^{2}_-}{\sigma^2_{peak}}\right)\right)~,\lb{bb}
\ee
where $\sigma_{-}\equiv \sigma_{H}(t_{k}\ll t_{k_{peak}})$ and 
$\sigma_{peak}\equiv \sigma_{H}(t_{k_{peak}})$.
Though $\frac{\sigma_{-}}{\sigma_{peak}}$ is an increasing function of $p$ -- we only consider 
here $p<1$ and larger $p$ corresponds to a {\it smaller} step -- the decrease of
$\sigma_{-}$ itself with increasing $p$ produces a faster decrease of the exponent, 
and hence a decrease of the right-hand side of (\ref{bb}). 
In other words, a smaller $p$, which corresponds to a larger step, increases 
the quantity $\frac{\beta_-}{\beta_{peak}}$.
On the other hand, as can be seen from (\ref{rapport}), $\frac{\beta_-}{\beta_{peak}}$ cannot 
grow arbitrarily. Hence, this results in an absolute lower bound $p_{\gamma}$ for the values of 
the parameter $p_{min}$ for which $\Omega_{PBH,0}(M_{peak})=0.3$ is satisfied, namely 
$p_{\gamma}\simeq 7.58\times 10^{-4}$, and a corresponding upper bound on $M_s$ at 
$5\times 10^{20}$g. For higher $M_s$, $p_{min}<p_{\gamma}$ 
is needed in order to achieve $\Omega_{PBH,0}=0.3$, and therefore  
the $\gamma$-ray background constraint is no longer satisfied. 
All these constraints are displayed in Fig.~\ref{constraint}.
Of course we obviously obtain a lower bound on $M_s$ around 
$4\times 10^{15}$g (for the same gravitational constraint) because we would violate again the
$\gamma$-ray constraint due to the presence of the bump close to the mass range $\Delta M$.
To summarize, requiring that $\Omega_{PBH,0}=0.3$, the $\gamma$-ray background constraint
can be satisfied for a range of characteristic masses $M_s$ with
\be
10M_{*}\lesssim M_s \lesssim 10^{21}\ {\rm g}~.\lb{Ms} 
\ee

An observational constraint on the allowed range of MACHO masses could therefore 
give a severe constraint on our model as it must correspond to the window (\ref{Ms}).
Though such observations do not exist for the moment, it is interesting that our model has this 
built-in constraint. In particular, it is interesting that our model can be easily discriminated 
from models where PBHs with $M\sim {\rm M}_{\odot}$ are produced.
It is instructive to study the dependence of the window (\ref{Ms}) on the present density of PBHs.
We can relax the gravitational constraint yielding $\Omega_{PBH,0}=0.3$
in order to allow a bigger upper bound on $M_s$ and to enlarge the window (\ref{Ms}).
Actually, we can consider the $\gamma$-ray constraint independently of the present 
density of PBHs, i.e., independently of the value $\beta_{peak}$. Obviously this constraint is given by 
\be
\beta_{-}\le 1.6\times 10^{-25} \left(\frac{M_{H}}{10^{15}{\rm g}}\right)^\frac{1}{2}~h^{2}\lb{be_}
\ee
with $M_{H}\in\Delta M$. 
The constraint (\ref{be_}) results in a value for $p_{\gamma}$ which is independent of $M_s$ 
as long as $M_s\gtrsim 2.5\times 10^{16}\ {\rm g}$. 
Hence for all those $M_s$ values, the minimum value $p_{\gamma}$
in order to satisfy the $\gamma$-ray constraint will be the same, namely the value mentioned 
earlier, $p_{\gamma}=7.58\times 10^{-4}$.
Thus, in the way shown in Fig.~\ref{constraint}, 
relaxing the requirement on the present density of PBHs, for example setting $\Omega_{PBH,0}=0.1$, 
allows a slightly weaker constraint for the upper bound on
$M_s\simeq 4\times 10^{21}\ {\rm g}$, 
and one gets a maximum $M_s\simeq 4\times 10^{23}\ {\rm g}$ by requiring only 
$\Omega_{PBH,0}=0.01$.
The lower bound on $M_s$ remains essentially the same around 
$3\times 10^{15}\ \rm g$.

We would like to make here a final comment. Recent numerical simulations seem to imply 
that PBH formation proceeds via near-critical collapse \cite{NJ98} whereby PBH of different masses 
around the horizon mass $M_H(t_k)=M_{peak}$ could be formed at the same time $t_k$, with 
\be
M = K ( \delta - \delta_c )^{\gamma}~.\lb{cc}
\ee
For $\gamma=0.35$, following \cite{Y98}, the following result is obtained, 
\be
\frac{d \Omega_{PBH}(M,t_{k_{peak}})}{d M}
\simeq 3.86~\frac{\beta(M_{peak})}{M_{peak}} 
          \left ( \frac{M}{M_{peak}} \right )^{2.86}~
\exp [ -1.35 \left (\frac{M}{M_{peak}} \right )^{2.86} ] ~.\lb{Y} 
\ee
This is turn yields 
\be
\Omega_{PBH,tot}(t_{k_{peak}}) = 0.8 ~\beta(M_{peak})~,\lb{omcc}
\ee
a result similar to that obtained in the more traditional view, where PBHs with 
different masses $M$ are produced at different times $t_k$ with $M=M_H(t_k)$; 
in the latter case one obtains $\Omega_{PBH,tot}(t_{k_{peak}}) = \beta_{peak}$. 
Hence, our constraint on $\Omega_{PBH,0}(M_{peak})$ using (\ref{gc3}) 
is stronger than the requirement 
$\Omega_{PBH,tot}(t_0)\leq 0.3$ based on (\ref{omcc}).
Note that $M_{peak}$ ($M_H$ in the notation of \cite{Y98}) is the Hubble mass corresponding 
to the peak in the quantity $\beta(M)$, not in the primordial 
spectrum itself, because one has to distinguish between the quantities 
$\sigma_H(t_k)$ and $\delta_H(t_k)$.
The above mentioned approach does not take into account the properties 
of the feature in the primordial spectrum, in particular the 
width of the corresponding bump in $\beta$, besides the value $\beta(M_{peak})$ which serves as a 
kind of overall normalization of the PBH abundance. 
On the other hand, it does account in a consistent way for the production of 
PBHs also in the monotonically increasing part of $\beta(M)$
around $M_{peak}$, $M\lesssim M_{peak}$. 
Clearly, the more spiky the primordial spectrum, the better this approach is 
expected to be, so our model provides such a concrete spectrum 
based on an underlying inflationary dynamics. 
An accurate determination of the mass spectrum would therefore yield valuable 
information, not only on the underlying primordial perturbations spectrum but 
also on PBH formation itself.

\section{Conclusions}
New expressions were derived recently for the improved computation of the mass variance at
horizon crossing \cite{BKP1}. We had found that for a scale-free powerlaw primordial spectrum, 
the mass variance for PBH masses around $M_*\approx 5\times 10^{14}\ {\rm g}$ is significantly
lower than it had been assumed in the literature.
In this paper we have refined our estimate and confirmed this discrepancy: We find that 
$5.37\leq \alpha^2(M_*)\leq 6.83$, for $1\leq n\leq 1.3$, see Eq.~(\ref{alphanum}), in particular 
$\sigma^2_H(t_k)= 6.63~\delta^2_H(t_0)$ for a scale-invariant 
(Harrison-Zel'dovich) primordial 
spectrum. For scale-free primordial spectra, the quantity $\alpha$ is further independent of $M$ 
for a wide range of scales. 
This is to be compared with the value $\sigma^2_H(t_k)\approx 25 ~\delta^2_H(t_0)$ which corresponds to 
the normalization adopted previously in the literature.

We have further applied our formalism to primordial spectra 
which are not scale-free. Two specific 
spectra possessing a characteristic scale were considered. The first one is purely 
phenomenological and corresponds to a pure step in the primordial spectrum. We find that the 
step is translated into the probability $\beta(M)$ in a non-trivial way: It acquires a smoothed step 
centered around the characteristic mass $M_s$. The second primordial spectrum corresponds 
to a BSI spectrum which is obtained from inflationary perturbations when the inflaton potential 
has a jump in its first derivative at some scale $k_s$.
It would of course be of interest to relate this scale to a scale
of fundamental physics at the GUT scale. 
In this case, we find that the mass variance $\sigma^2_H(t_k)$ 
displays a sharp, pronounced peak centered at the scale $k_{peak}\approx 1.27 ~k_s$ 
which is {\it not} the exact location of the maximal peak in the primordial spectrum itself. 
Again this is due to the non-trivial relationship between the primordial perturbations 
spectrum itself and the mass variance at Hubble radius crossing $\sigma^2_H(t_k)$.
This in turn yields a pronounced bump in the quantity $\beta(M)$ around $M_{peak}$ 
which has interesting applications. We have considered here in detail one such application 
\cite{BKP2}, the possibility to produce in this model a significant part of the cold dark 
matter in our Universe. Interestingly, we find that the mass of the produced PBHs is constrained 
to be much smaller than one solar mass. Indeed, the characteristic mass of the spectrum, 
essentially the same as that of the produced PBHs, cannot be fixed arbitrarily because of the 
severe constraint coming from the $\gamma$-ray background. 
The inclusion of a non-negligible spectrum of primordial gravitational waves and 
running scalar and tensorial spectral indices $n_s$ and $n_T$ (see e.g.\cite{DS01}) 
could enlarge the class of models that can realize our scenario.
A thorough study should probably await a more careful determination of $\delta_{min}$ and, 
to less extent, of the cosmological parameter $\Omega_{\Lambda}$.
At the hand of the example studied here, it is 
clear that spectra with characteristic scales on very small scales, much smaller than those 
probed by the CMB anisotropy or large-scale structures, offer 
interesting new avenues.    

We have studied here concrete models with concrete assumptions,
 for which we have obtained precise quantitative results. It is clear that
these results may be different for other models. They may also depend on
the assumptions made. In particular, the influence of non-Gaussian
fluctuations could strongly change the PBH formation rate
\cite{BP}. However, all situations can in principle be dealt with
using the formalism presented here.

\end{document}